\documentclass{article}                    
\usepackage{graphicx}
\usepackage{geometry}
\usepackage{cite}

\begin{document}

\title{\bf{The contribution of Giordano Bruno}\\{\bf to the  principle of relativity}}


\author{Alessandro De Angelis\\
              INFN - Istituto Nazionale di Fisica Nucleare, sezione di Padova 
                            and INAF  Padova \vspace{1 mm} \\ 
                         Catarina Espirito Santo\\
           LIP - Laborat\'orio de Instrumenta\c{c}\~ao e F\'isica Experimental de Part\'iculas, Lisboa}



\maketitle

\begin{abstract}
The trial and condemnation of Giordano Bruno was mainly based on arguments of philosophical and theological nature, and therefore different from Galilei's. Such elements contribute to unfairly devalue the scientific contribution of Bruno and do not properly account in particular for his contribution to physics.  This paper discusses the contribution of Bruno to the principle of relativity. According to common knowledge, the special principle of relativity was first enunciated in 1632 by Galileo Galilei in his {\em Dialogo sopra i due massimi sistemi del mondo (Dialogue concerning the two chief world systems),} using the metaphor today known as ``Galileo's ship'': in a boat moving at constant speed, the mechanical phenomena can be described by the same laws holding on Earth. We shall show that the same metaphor and some of the examples in Galilei's book were already contained in the dialogue {\em La cena de le Ceneri (The Ash Wednesday Supper)} published by Giordano Bruno in 1584. In fact, Giordano Bruno largely anticipated the arguments of Galilei on the relativity principle, in particular to support the Copernican view. It is likely that Galilei was aware of Bruno's work, and it is possible that the young Galilei discussed with Bruno, since they both stayed in Venezia for long periods in 1592.
\vskip 2mm


\vskip 2mm
\noindent PACS{01.65.+g}
\end{abstract}

\vskip 4 cm
\begin{center} (To be published in Journal of Astronomical History and Heritage) \end{center}

\newpage

\section{Introduction}
\label{intro}

The principle of relativity states that it is impossible to determine whether a system is at rest
or moving at constant speed with respect to an inertial system  by experiments internal to the system, i.e.,
there is no internal observation by which one can distinguish a system moving uniformly from one at rest.
This principle played a key role in the defence of the heliocentric system, 
as it made the movement of the Earth compatible with everyday experience.

According to common knowledge, the principle of relativity was first enunciated by Galileo Galilei (1564 -- 1642) 
in 1632 in his {\it Dialogo sopra i due massimi sistemi del mondo} 
({\it Dialogue concerning the two chief world systems}) \cite{galilei}, 
using the metaphor known as ``Galileo's ship'': 
in a boat moving at constant speed, the mechanical phenomena can be described by the same laws holding on Earth.

Many historical aspects of the birth of the relativity principle have received little or scattered attention. 
In this short note, we put together some elements 
showing that Giordano Bruno largely anticipated \cite{bruno} the arguments of Galilei on the relativity principle.
In addition, we briefly discuss the silence of Galilei on Bruno and  the relation
between the lives and careers of the two scientists. 

\section{Galilei and the principle of relativity}
\label{galilei}

The {\it Dialogo sopra i due massimi sistemi del mondo} 
is the source usually quoted  for the enunciation of the principle of relativity by Galilei. 
However, its publication in 1632 was certainly not a surprise, as Galilei had expressed his views much before, in particular when lecturing
 at the University of Padova from 1592 to 1610. 
Some aspects of the evolution of Galilei's ideas, from the {\it Trattato della sfera} (1592) \cite{sfera} in which the Earth is still placed at the centre of the Universe,
towards the {\it Dialogo}, passing through his heliocentric correspondence with Kepler from 1597 onwards \cite{lettere},
are  examined for example in \cite{martins,giannetto,7,8,9,10,clavelin}.

In February 1616, the Roman Inquisition condemned the theory by Nicolaus Copernicus (1473 -- 1543)  \cite{copernico} as being ``foolish and absurd in philosophy''.
The month before, the inquisitor Monsignor Francesco Ingoli addressed Galilei with the essay 
{\it Disputation concerning the location and rest of Earth against the system of Copernicus} \cite{ingoli}. 
The letter listed both scientific and theological arguments against Copernicanism. 
Galilei replied only in 1624. In his lengthy reply, he introduced an early version of ``Galileo's ship'' metaphor, and discussed the experiment of dropping a stone from the
top of the mast. 
Both arguments, as we shall see, had been previously raised by Bruno, and will be later used again by Galilei, 
although with small differences,
in the {\it Dialogo}. 

In the {\it Dialogo sopra i due massimi sistemi del mondo}, 
Galilei discusses  the arguments then current against the idea that the Earth moves.
The book is a fictional dialogue between three characters. Two of these, Salviati and Sagredo, refer to real figures 
disappeared since a few years at the publication of the book. 
The first plays the role of the defender of the Copernican theory, putting forward Galilei's point of view; 
the second, a Venetian aristocrat, educated and liberal, willing to accept new ideas, acts as a moderator
placed between Salviati and the third character, a certain Simplicio, fierce supporter of Aristotle. 
The name of the latter (reminiscent of ``simple-minded'' in Italian)
is in itself a clear indication of the Galilean dialectics, designed to destroy the opponents. 
Simplicio, despite being one of the most famous commentators of Aristotle, 
manifests himself with an embarrassing simplicity of spirit. 
Galilei uses Salviati and Simplicio as spokespersons of the two clashing chief world systems; 
Sagredo represents the discreet reader, the steward of science, the one to whom the book is addressed: 
he intervenes in the discussions asking for clarifications, contributing conversational topics, acting like a science enthusiast.

In the second day, Galilei's dialogue considered Ingoli's arguments against the idea that the Earth moves. 
One of these is that if the Earth were spinning on its axis, then we would all be moving eastward at hundreds of miles per hour, 
so a ball dropped down from a tower would land west of the tower that would have in the meantime moved a certain distance eastwards. 
Similarly, the argument went, a cannonball shot eastwards would fall closer to the cannon compared to a ball shot westwards, 
since the cannon moving East would partly catch up with the ball. 

To counter such arguments Galilei proposes through the words of Salviati a {\em gedankenexperiment}: 
examine the laws of mechanics in a ship moving at a constant speed. 
Salviati claims that there is no internal observation which allows to distinguish between a system smoothly moving and one 
at rest. So two systems moving without acceleration are equivalent, and non-accelerated motion is relative:

\vspace{2mm}
{\em ``Salviati -- Shut yourself up with some friend in the main cabin below decks on some large ship, and have with you there 
some flies, butterflies, and other small flying animals. Have a large bowl of water with some fish in it; 
hang up a bottle that empties drop by drop into a wide vessel beneath it. 
With the ship standing still, observe carefully how the little animals fly with equal speed to all sides of the cabin. 
The fish swim indifferently in all directions; the drops fall into the vessel beneath; 
and, in throwing something to your friend, you need throw it no more strongly in one direction than another, the distances being equal; 
jumping with your feet together, you pass equal spaces in every direction. 
When you have observed all these things carefully (though doubtless when the ship is standing still everything must happen in this way), 
have the ship proceed with any speed you like, so long as the motion is uniform and not fluctuating this way and that. 
You will discover not the least change in all the effects named, 
nor could you tell from any of them whether the ship was moving or standing still. 
In jumping, you will pass on the floor the same spaces as before, 
nor will you make larger jumps toward the stern than toward the prow even though the ship is moving quite rapidly, 
despite the fact that during the time that you are in the air the floor under you will be going in a direction opposite to your jump. 
In throwing something to your companion, you will need no more force to get it to him 
whether he is in the direction of the bow or the stern, with yourself situated opposite. 
The droplets will fall as before into the vessel beneath without dropping toward the stern, 
although while the drops are in the air the ship runs many spans. 
The fish in their water will swim toward the front of their bowl with no more effort than toward the back, 
and will go with equal ease to bait placed anywhere around the edges of the bowl. 
Finally the butterflies and flies will continue their flights indifferently toward every side, 
nor will it ever happen that they are concentrated toward the stern, as if tired out from keeping up with the course of the ship, 
from which they will have been separated during long intervals by keeping themselves in the air. 
And if smoke is made by burning some incense, it will be seen going up in the form of a little cloud, 
remaining still and moving no more toward one side than the other. 
The cause of all these correspondences of effects is the fact that the ship's motion is common to all the things contained in it, 
and to the air also. That is why I said you should be below decks; for if this took place above in the open air, 
which would not follow the course of the ship, more or less noticeable differences would be seen in some of the effects noted.'' 
}

\vspace{2mm}
Note that Galilei does not state that the Earth is moving, 
but that the motion of the Earth and the motion of the Sun cannot be distinguished (hence the name of relativity):

\vspace{2mm}
{\em ``There is one motion which is most general and supreme over all, and it is that by which the Sun, Moon, 
and all other planets and fixed stars -- in a word, the whole universe, the Earth alone excepted -- appear to be moved as a unit 
from East to West in the space of twenty-four hours. 
This, in so far as first appearances are concerned, may just as logically belong to the Earth alone as to the rest of the Universe, 
since the same appearances would prevail as much in the one situation as in the other.''}

\section{Relativity and celestial motion before Copernicus}

The possibility that the Earth moves had been discussed several times in particular by the Greeks, mostly as a hypothesis to be rejected. Also an annual motion of the Earth around the Sun had been considered, by Aristarchus of Samos (c. 310 -- c. 230 BC). Unfortunately, very little is known about Greek astronomy as a science, being many of our reconstructions a speculation. Later, some medieval authors discussed the possibility of the Earth's daily rotation. The first notable example is probably Jean Buridan (c. 1300 -- 1361), one of the ``doctores parisienses'' -- a group of professors at the University of Paris in the fourteenth century, including notably Nicole Oresme. 

In Buridan the example of the ship, later used in particular by Oresme, Bruno, and Galilei, is contained in the Book 2 of his commentary \cite{14} to Aristotle's {\em{On the heavens}} 
\cite{aristo}: {\em ``It should be known that many people have held as probable that it is not contradictory to appearances for the Earth to be moved circularly in the aforesaid manner, and that on any given natural day it makes a complete rotation from west to east by returning again to the west - that is, if some part of the Earth were designated [as the part to observe]. Then it is necessary to posit that the stellar sphere would be at rest, and then night and day would result through such a motion of the Earth, so that motion of the Earth would be a diurnal motion. The following is an example of this: if anyone is moved in a ship and imagines that he is at rest, then, should he see another ship which is truly at rest, it will appear to him that the other ship is moved. This is so because his eye would be completely in the same relationship to the other ship regardless of whether his own ship is at rest and the other moved, or the contrary situation prevailed. And so we also posit that the sphere of the Sun is totally at rest and the Earth in carrying us would be rotated. Since, however, we imagine we are at rest, just as the man on the ship moving swiftly does not perceive his own motion nor that of the ship, then it is certain that the Sun would appear to us to rise and set, just as it does when it is moved and we are at rest.''}

It appears to us, in agreement with Barbour \cite{10}, that the relativity invoked by Buridan is in the first place kinematic. In the words of Barbour, {\em ``we have [here] a clear statement of the principle of relativity, certainly not the first in the history of the natural philosophy of motion but perhaps expressed with more cogency than ever before. The problem of motion is beginning to become acute. We must ask ourselves: is the relativity to which Buridan refers kinematic relativity or Galilean relativity? There is no doubt that it is in the first place kinematic; for Buridan is clearly concerned with the conditions under which motion of one particular body can be deduced by observation of other bodies.''}

Buridan writes later on, still in \cite{14}: {\em ``But the last appearance which Aristotle notes is more demonstrative in the question at hand. This is that an arrow projected from a bow directly upward falls to the same spot on the Earth from which it was projected. This would not be so if the Earth were moved with such velocity. Rather, before the arrow falls, the part of the Earth from which the arrow was projected would be a league's distance away. But still supporters would respond that it happens so because the air that is moved with the Earth carries the arrow, although the arrow appears to us to be moved simply in a straight line motion because it is being carried along with us. Therefore, we do not perceive that motion by which it is carried with the air.''} Some worry about dynamics is, thus, already in Buridan. But the conclusion is that {\em ``the violent impetus of the arrow in ascending would resist the lateral motion of the air so that it would not be moved as much as the air. This is similar to the occasion when the air is moved by a high wind. For then an arrow projected upward is not moved as much laterally as the wind is moved, although it would be moved somewhat.''} The theory of impetus is thus not pushed to the limit in which one could identify it with the principle of inertia, nor with a dynamical concept of relativity.

A further step is implicitly made, few years later, by Nicole Oresme (c. 1320 -- 1382) in \cite{oresme}. Oresme first states that no observation can disprove that the Earth is moving: {\em ``[O]ne could not demonstrate the contrary by any experience. [...] I assume that local motion can be sensibly perceived only if one body appears to have a different position with respect to another. And thus, if a man is in a ship called a which moves very smoothly, irrespective if rapidly or slowly, and this man sees nothing except another ship called b, moving exactly in the same way as the boat a in which he is, I say that it will seem to this person that neither ship is moving.''}

Oresme also provides an argument against Buridan's interpretation of the example of the arrow (or stone in the original by Aristotle) thrown upwards, introducing the principle of composition of movements: {\em ``[O]ne might say that the arrow thrown upwards is moved eastward very swiftly with the air through which it passes, with all the mass of the lower part of the world mentioned above, which moves with a diurnal movement; and for this reason the arrow falls back to the place on the Earth from which it left. And this appears possible by analogy, since if a man were on a ship moving eastwards very swiftly without being aware of his movement, and he drew his hand downwards, describing a straight line along the mast of the ship, it would seem to him that his hand was moved straight down. Following this opinion, it seems to us that the same applies to the arrow moving straight down or straight up. Inside the ship moving in this way, one can have horizontal, oblique, straight up, straight down, and any kind of movement, and all look like if the ship were at rest. And if a man walks westwards in the boat slower than the boat is moving eastwards, it will seem to him that he is moving west while he is going east.''}

Also Nicolaus Cusanus (1401 -- 1461) stated later, without going into detail, that the motion of a ship could not be distinguished from rest on the basis of experience, but some different arguments need to be invoked -- and the same applies {\em ``to the Earth, the Sun, or another star''} \cite{17}.

All this happens before Copernicus: we still discuss how things could be, not much how things ``are''. This view will change after Copernicus.

\section{Giordano Bruno and the principle of relativity}
\label{bruno}

In April 1583, forty years after the publication of the book by Copernicus and nine years before the then 28-years old Galilei was called to the University of Padova, 
Bruno went to England and lectured in Oxford, unsuccessfully looking for a teaching position there. Still, 
the English period was a fruitful one. During that time Bruno completed and published some of his most important works, 
the six ``Italian Dialogues", including the cosmological work {\it La cena de le Ceneri} ({\it The Ash Wednesday Supper}, 1584)
\cite{bruno}.

This book consists of five dialogues between Theophilus, a disciple that exposes the theories of Bruno; 
Smitho, a character probably real but difficult to identify, possibly an English friend of Bruno 
(perhaps John Smith or the poet William Smith) -- the Englishmen has simple arguments, but he has good common sense and is free of prejudice;
Prudencio, a pedantic character, and Frulla, also a fictional character who, as the name in Italian suggests, embodies a comic figure, 
provocative and somewhat tedious with the proposition of stupid arguments. 

In the third dialogue, 
the four mostly comment discussions heard at a supper attended by Theophilus in which Bruno -- called in the text ``il Nolano'' (the Nolan), 
because he was born in Nola near Naples -- was arguing in particular with Dr. Torquato and Dr. Nundinio, 
representing the Oxonian faculty.
Bruno starts by discussing the argument related to the air, winds and the movement of clouds. He largely uses the fact that the air is dragged by the Earth:

\vspace{2mm}
{\em 
``Theophilus --  [...] If the Earth were carried in the direction called East, it would be necessary that the clouds in the air should always appear moving toward west, because of the extremely rapid and fast motion of that globe, which in the span of twenty-four hours must complete such a great revolution. To that the Nolan replied that this air through which the clouds and winds move are parts of the Earth, because he wants (as the proposition demands) to mean under the name of Earth the whole machinery and the entire animated part, which consists of dissimilar parts; so that the rivers, the rocks, the seas, the whole vaporous and turbulent air, which is enclosed within the highest mountains, should belong to the Earth as its members, just as the air does in the lungs and in other cavities of animals by which they breathe, widen their arteries, and other similar effects necessary for life are performed. The clouds, too, move through happenings in the body of the Earth and are based in its bowels as are the waters. [...] Perhaps this is what Plato meant when he said that we inhabit the concavities and obscure parts of the Earth, and that we have the same relation with respect to animals that live above the Earth, as do in respect to us the fish that live in thicker humidity. This means that in a way the vaporous air is water, and that the pure air which contains the happier animals is above the Earth, where, just as this Amphitrit\footnote{Amphitrite was in Greek mythology the wife of Poseidon and, therefore, the goddess of  sea.} [ocean] is water for us, this air of ours is water for them. This is how one may respond to the argument referred to by Nundinio; just as the sea is not on the surface, but in the bowels of the Earth, and just as the liver, this source of fluids, is within us, that turbulent air is not outside, but is as if it were in the lungs of animals.''
}

\vspace{2mm}
The Dialogue than moves to discussing the argument of projectiles. He starts by explaining the Aristotelian objection of the
stone sent upwards:

\vspace{2mm}
{\em 
``Smitho -- You have satisfied me most sufficiently, and you have excellently opened many secrets of nature which lay hidden under 
that key. Thus, you have replied to the argument taken from winds and clouds; 
there remains yet the reply to the other argument which Aristotle submitted in the second book of 
{\em{On the Heavens}}\footnote{See \cite{aristo}, Section 296b.} 
where he states that it would be impossible that a stone thrown high up could come down along the same perpendicular straight line, 
but that it would be necessary that the exceedingly fast motion of the Earth should leave it far behind toward the West. 
Therefore, given this projection back onto the Earth, it is necessary that with its motion there should come a change in all relations of 
straightness and obliquity; just as there is a difference between the motion of the ship and the motion of those things that arc on the ship 
which if not true it would follow that when the ship moves across the sea one could never draw something along a straight line from one of 
its corners to the other, and that it would not be possible for one to make a jump and return with his feet to the point from where 
he took off.''
}

\begin{figure}
\center{\includegraphics[width=0.6\columnwidth]{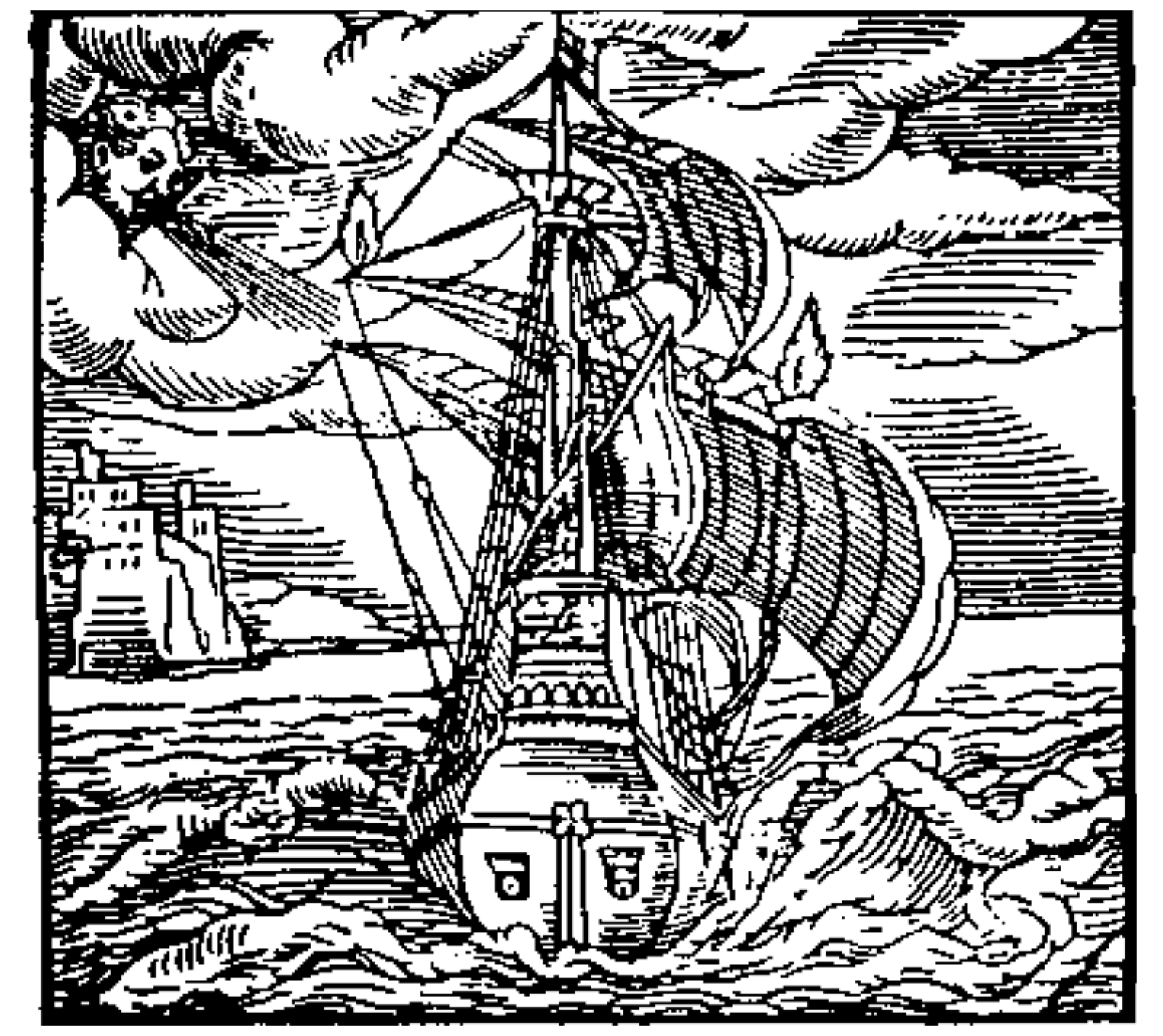}}
\caption{The figure of the ship referred in the text of the dialogue; note that it lacks the lettering referred to.\label{ship}}
\end{figure}

\vspace{2mm}
Bruno then gives, in Theophilus's speech, the following reply (referring to Figure \ref{ship} in the text):

\vspace{2mm}
{\em
``Theophilus -- With the Earth move [...] all things that are on the Earth.
If, therefore, from a point outside the Earth something were thrown upon the Earth, it would lose, because of the latter's motion, 
its straightness as would be seen on the ship AB moving along a river, if someone on point C of the riverbank were to throw a stone along 
a straight line, and would see the stone miss its  target by the amount of the velocity of the ship's motion.  
But if someone were placed high on the mast of that ship, move as it may however fast, he would not miss his target at all, 
so that the stone or some other heavy thing thrown downward would not come along a straight line from the point E which is at the top of the mast, or cage, to the point D which is at the bottom of the mast, or at some point in the bowels and body of the ship. 
Thus, if from the point D to the point E someone who is inside the ship would throw a stone straight up, 
it would return to the bottom along the same line however far the ship moved, provided it was not subject to any pitch and roll.''
}

\vspace{2mm}
He then continues with the statement that the movement of the ship is irrelevant for the events occurring 
within the ship and explains the reasons for that:

\vspace{2mm}
{\em
``[...] If there are two, of which one is inside the ship that moves and the other outside it, 
of which both one and the other have their hands at the same point of the air, 
and if at the same place and time one and the other let a stone fall without giving it any push, 
the stone of the former would, without a moment's loss and without deviating from its path, go to the prefixed place, 
and that of the second would find itself carried backward. 
This is due to nothing else except to the fact that the stone which leaves the hand of the one supported by the ship, 
and consequently moves with its motion, has such an impressed virtue, which is not had by the other who is outside the ship, 
because the stones have the same gravity, the same intervening air, if they depart (if this is possible) from the same point, 
and arc given the same thrust.

From that difference we cannot draw any other explanation except that the things which are affixed to the ship, and belong to it 
in some such way, move with it: and the stone carries with itself the virtue [impetus] of the mover which moves with the ship. 
The other does not have the said participation. From this it can evidently be seen that the ability to go straight comes not from the point 
of motion where one starts, nor from the point where one ends, nor from the medium through which one moves, 
but from the efficiency of the originally impressed virtue, on which depends the whole difference. 
And it seems to me that enough consideration was given to the propositions of Nundinio.''
}

\vspace{2mm}
The experiments carried out in a ship are thus not influenced by its movement
because all the bodies in the ship take part in that movement, regardless of whether they are in contact with the ship or not.
This is due to the ``virtue'' they have, which remains during the motion, after the carrier abandons them. Bruno thus clearly expresses the concept of inertia, using the word ``virt\`u",
in Italian meaning ``quality'', which is carried by the bodies moving with the ship -- and with the Earth. The arguments of Bruno certainly constitute a significant step towards the principle of inertia.

\section{Discussion}
\label{disc}

We have seen that Bruno in La cena de le Ceneri anticipates to a great extent the arguments of Galilei on the principle of relativity. In fact, his explanation does contain all the fundamental elements of the principle. The idea that the only movement observable by the subject is the one in which he does not take part, was present before in the works of Buridan, and later by Nicole Oresme together with the notion of the composition of movements, alien to Aristotelian mechanics [9]. Similar arguments were used by Copernicus \cite{copernico}. The main missing ingredient was the idea of inertia, which explains the fact that projectiles move along with Earth. In fact, while there is a continuous line between Buridan, Oresme, Copernicus, Bruno, and Galilei, the arguments of Bruno on the impossibility to detect absolute motion by phenomena in the ship constitute a significant step towards the principle of inertia and to a dynamical context for relativity. What is new in Bruno, and what brings him almost exactly to where Galilei stood, is thus a clear concept on inertia.

The arguments and metaphors used in discussions concerning the world systems were common to different authors. They were largely derived from Aristotle, Ptolemy and their commentators, often used without referencing and sometimes attributed to the wrong source. For example, Aristotle in his On the heavens uses as experimental argument the one of the stone sent upwards. Buridan and Oresme, in their comment to this work, used a modified version of this experiment in which an arrow is sent upwards in a ship - probably introduced by an earlier commentator/translator. Nevertheless, the description by Galilei of exactly the same ship experiment that Bruno used in the Cena makes it very likely that Galilei knew this work. The use of the dialogue form with a similar choice of characters can also be seen as a possible sign of the influence of Bruno in Galilei.

On the other hand, Galilei never mentioned Bruno in his works, and in particular there is no reference to him in Galilei's large corpus of letters, while he referenced the ``doctores parisienses'' in his MS 46 [18], a 110 pages long manuscript, containing physical speculations based on Aristotle's {\em{On the Heavens}}.  
Some authors have commented on Galilei's silence about Bruno putting forward reasons of prudence \cite{clavelin}.
As pointed out in \cite{martins}, this can hardly explain the absence of any mention also in his personal correspondence.
Furthermore, although Galilei himself never mentions the name of Bruno in his personal notes and letters,
several of his correspondents do mention the Nolan.
Martin Hasdale, in a letter to Galilei from 1610, tells him that Kepler expressed his admiration for 
Galilei, although he regretted his failure to mention in his works Copernicus, Giordano Bruno and 
several Germans who had anticipated such discoveries -- in particular Kepler himself\footnote{Letter from M. Hasdale to G. Galilei \cite{lettere}.}. 

\vspace{2mm}
{\em
``This morning I had the opportunity to make friends with Kepler [...] I
asked what he likes about that book of yourself and he replied that
since many years he exchanges letters with you, and that he is really
convinced that he does not know anybody better than you in this
profession. [...] As for this book, he says that you really showed the
divinity of your genius; but he was somehow uneasy, not only for the
German nation, but also for your own, since you did not mention those
authors who introduced the subject and gave you the opportunity to
investigate what you found now, naming among these Giordano Bruno
among the Italians, and Copernicus, and himself."
}
\vspace{2mm}

We can thus say that Galileo Galilei was probably aware of the work by Giordano Bruno on the Copernican system. 
When Galilei arrived in Padova, in 1592, it is also possible that the two scientists met:  
Bruno was guest of the noble Giovanni Mocenigo in Venezia, and Galilei was living between Padova and Venezia. 
In 1591, Bruno had unsuccessfully applied for the chair of Mathematics that was assigned to Galilei one year later. 
Although a confirmation that such a meeting occurred might be difficult to find, it remains hard to believe, 
given the motivations and characters of the two men and the circumstances of their lives during those years, 
as well as the size of the literate and scientific community in those days, that they failed to discuss on the 
arguments concerning the defence of the Copernican system.

\paragraph{Acknowledgenents --} We thank Luisa Bonolis, Alessandro Bettini, Alessandro Pascolini, Giulio Peruzzi and Antonio Saggion for useful suggestions. We thank the anonymous referee for indicating us some important aspects that we neglected in the first draft.


\end{document}